\documentclass[12pt]{article}
\usepackage{graphicx}
\usepackage[font=small,labelfont=bf]{caption}
\usepackage{authblk}
\usepackage{natbib}
\title{\huge Multi-channel free-space optical convolutions with incoherent light}
\fontsize{12}{12}
\author[1,2]{\normalsize Alexander Song}
\author[1,2]{\normalsize Sai Nikhilesh Murty Kottapalli}
\author[1,2]{Peer Fischer}
\affil[1]{Max Planck Institute for Medical Research, Jahnstraße 29, 69120 Heidelberg, Germany}
\affil[2]{Institute for Molecular Systems Engineering and Advanced Materials, Universität Heidelberg, Neuenheimer Feld 225, 69120 Heidelberg, Germany}
\date{}
\begin{document}
\maketitle

\newpage

\section{Abstract} 
Free-space optical systems are promising candidates for high performance computing and have been particularly successful in the implementation of large-scale convolutions. Convolutions are the key operation in convolutional layers, which are used extensively in modern neural networks, especially in the context of image/video processing and generation. These optical accelerators have demonstrated remarkable performance in both processing rates and energy efficiency. Prior approaches have primarily demonstrated convolutions from a single input channel to one or more output channels. We extend these methods to perform true multi-channel convolutions, where multiple input channels are convolved with their own sets of convolutional kernels onto output channels. We simulate this approach using both ray-tracing and angular spectrum propagation and find the approach is highly-scalable. We then experimentally implement a proof-of-concept prototype to demonstrate multi-channel free-space optical convolutions.

\newpage

\section{Introduction}
Convolutional neural networks (CNNs) are widely used in machine learning applications such as image classification, object detection, and natural language processing \cite{gu2018recent}. Convolutional layers generate feature representations by convolving multiple input channels with learned kernels and summing the results into multiple output channels. Although the number of learnable parameters per kernel is moderate, modern networks employ large input sizes and hundreds of channels per layer, resulting in substantial computational cost, memory access, and energy consumption. These demands have motivated the exploration of alternative hardware platforms for accelerating convolutional operations.

Optical computing approaches are promising candidates for implementing convolution \cite{chang2018hybrid, feldmann2021parallel, shi2022loen, xu2022multichannel, hu2022high, wang2022optical, ju2023scalable, zheng2024multichannel, song2024low}, as light propagation naturally supports parallel linear operations while preserving spatial information. In particular, free-space optical systems are well suited for large-scale convolutions and for computation in general \cite{hu2024diffractive}. The Fourier transforming property of lenses enables convolution through spatial filtering in the Fourier plane \cite{chang2018hybrid, hu2022high}. Other approaches include lens-free convolution \cite{shi2022loen} and meta-optics-enabled real-valued convolution \cite{zheng2024multichannel}.

However, existing free-space implementations have primarily demonstrated single input channel to multiple output channel convolution. In practice, each output channel of a convolutional layer is formed by summing convolutions across multiple input channels \cite{aloysius2017review}. A straightforward extension of current optical systems requires sequential processing of input-channel and kernel sets followed by digital summation of the outputs. This hybrid strategy introduces significant data transfer and memory overhead, limiting throughput and reducing the advantages of optical parallelism.

\begin{figure}[ht]
\centering
\includegraphics[width=7.5cm,clip]{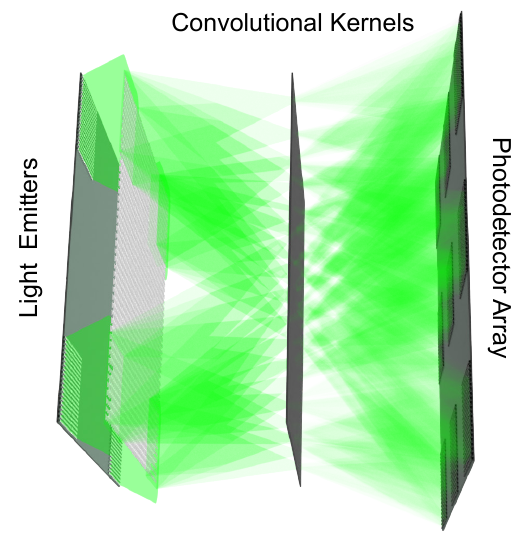}
\caption{3D render of optical multi-channel convolution implementation. Input channels are encoded through the intensity of arrays incoherent light emitters, such as an LED display (left). The emitted light is distributed by a microlens array onto a transmissive amplitude mask encoding convolutional kernels (center). The modulated light continues to propagate onto a photodiode array or camera (right) for readout.}
\label{fig-1} 
\end{figure}

In this work, we develop and experimentally demonstrate a free-space optical convolution system that performs multi-input, multi-output channel convolution in a single shot (Figure \ref{fig-1}). Arrays of incoherent light emitters encode the input channels, and a microlens array geometrically maps each channel onto convolutional kernels implemented as spatially varying weights on an amplitude mask. The convolved signals are spatially integrated into distinct output channel regions without electronic recombination. We validate the approach using reduced-resolution images of MNIST handwritten digits and analyze system scalability using both ray-tracing and Fourier optics simulations.

\section{Free-space multi-channel convolution concept}
\begin{figure}[ht]
\centering
\includegraphics[width=7.5cm,clip]{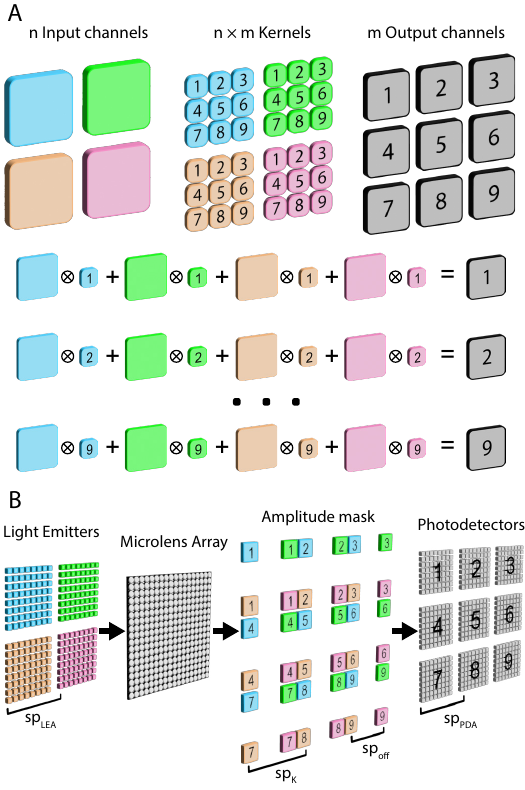}
\caption{(a) Concept of multichannel convolutions. In this example $n=4$ input channels, denoted by color, are mapped to $m=9$ output channels, which are numbered. The mapping is performed by a set of $n \times m$ convolutional kernels, one for each pair of input and output channels. Each output channel is calculated by convolving each input channel with the corresponding convolutional kernel and summing the results. (b) Schematic of optical multichannel convolutions. Light emitter arrays are used to encode input channels and are routed by a microlens array onto an amplitude mask, which encodes convolutional kernels on spatially non-overlapping positions. The convolved light is integrated onto separate output channels implemented with photodetector arrays.}
\label{fig-2} 
\end{figure}

Convolutional neural networks use convolutional layers to detect and extract local spatial patterns (like edges, textures, and shapes) from spatially structured data, such as images, by convolving filters across the input. These layers implement multi-channel convolutions (Figure \ref{fig-2}), where the input $I$ is described as a tensor of size $n \times H \times W$ and the output $O$ is a tensor of size $m \times H \times W$. Here, $n$ and $m$ denote the number of input and output channels, respectively, and $H \times W$ describes the size of the image.

The input and output tensors are related by convolutional filters, or kernels $K$, of size $k \times k$ for each pair $n,m$, resulting in a total tensor size of $k \times k \times n \times m$. These filters represent the total number of parameters to fit in the convolutional layer and remain moderate even for large numbers of input and output channels. Each output channel is defined as:
$$ O_{j=1...m} = \sum_{i=1}^{n} I_{i} \ast K_{ij} $$ 
where $K_{ij}$ is the corresponding kernel for a given input channel $i$ and output channel $j$. 

In our approach, we perform multi-channel convolutions using free-space optics (Figure \ref{fig-1}, \ref{fig-2}b). Input channels are encoded using the intensity of small, incoherent light emitters (LE), such as light-emitting diodes (LEDs). The LEs are arranged into $n$ subarrays, with each subarray corresponding to a single input channel with $H \times W$. The $n$ subarrays are tiled compactly across a 2D plane with approximately $\sqrt{n}$ subarrays to a side. 

Light from the emitters is mapped using a microlens array (MLA) and an amplitude mask with encoded convolutional kernels onto output channels. The position of the convolutional kernels depends on the geometry of the input and output channels, where two different grid spacings are each matched to the input and output channel positions and spacing. In this work, we use a liquid crystal display (LCD) to implement the amplitude mask. The weights on an LCD can be dynamically updated to display different convolutional kernels and consume low power for fixed weights. The light passing through the kernels continue to propagate onto the corresponding regions of the sensor.

\begin{figure}[ht]
\centering
\includegraphics[width=9cm,clip]{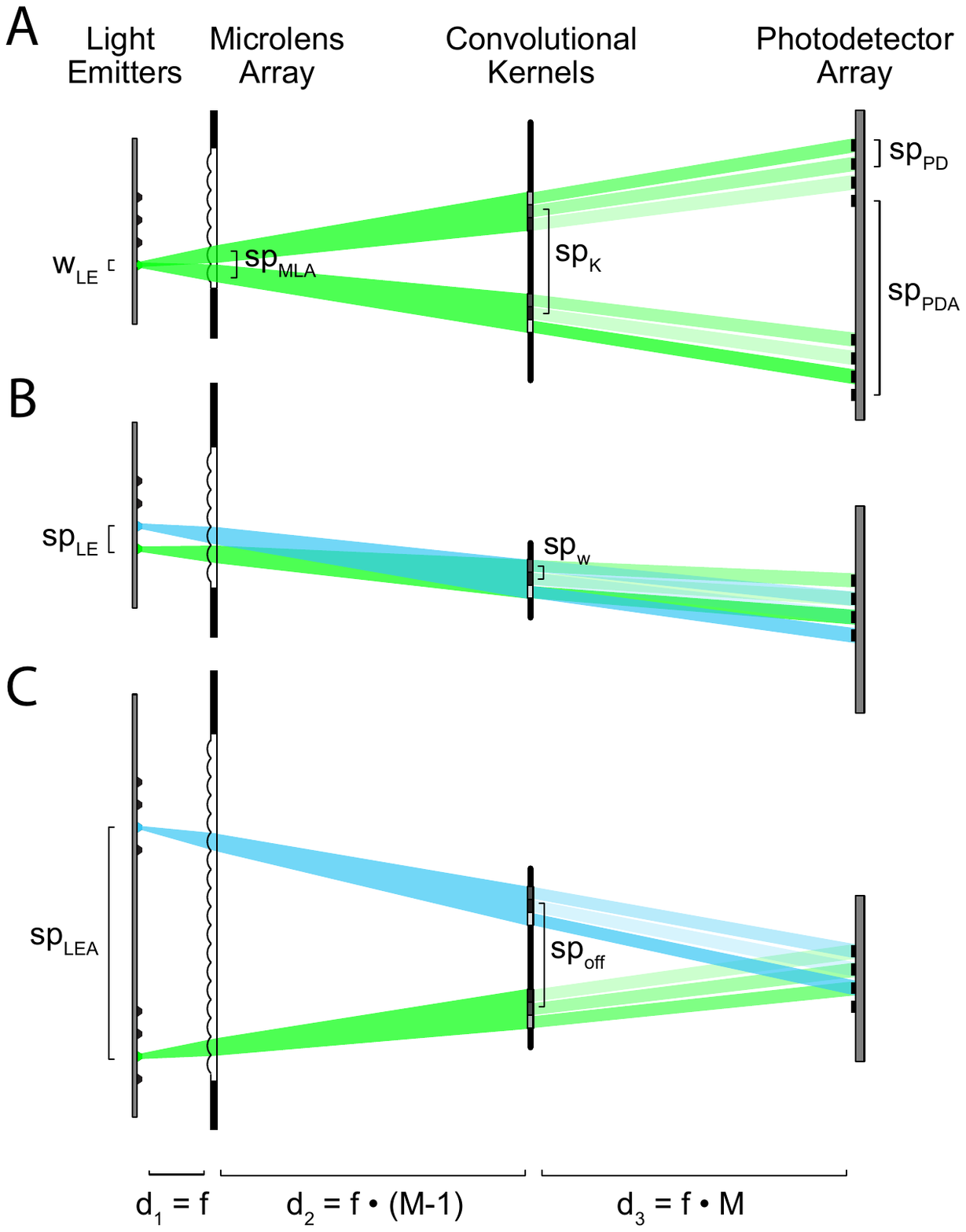}
\caption{Ray tracing diagram showcasing multi-kernel optical convolutions. (a) A light emitter is spread by the microlens array onto a set of kernels arrayed in 2D, corresponding to each of the $m$ output channels, before continuing to propagate onto the sensor array. (b) Adjacent light emitters on the same input channel (green, blue) interact with the same kernel before impinging on adjacent pixels of the sensor. (c) Light emitters from two different input channels (green, blue) are modulated by separate kernels on the amplitude mask before propagating onto the same output channel}
\label{fig-3} 
\end{figure}

The output channels are implemented with photodetectors (PDs) arranged in 2D, such as a photodiode array or a camera sensor. Similar to the input channels, the PDs are arranged in $m$ subarrays with each subarray corresponding to a single output channel with size $H \times W$. The $m$ subarrays are also tiled across a 2D plane with approximately $\sqrt{m}$ subarrays to a side. The different distances and sizes of optical components along with the propagation distances are interconnected and these relationships are summarized in Supplementary Table 1 and are depicted schematically in Figure \ref{fig-3}.

The light from a single LE (Figure \ref{fig-3}a) with width $w_{LE}$ first propagates a distance $d_1=f$, the focal length of a microlens. The MLA has a spacing of $sp_{MLA}$ and separates the light into collimated beamlets that continue to propagate a distance $d_2=f\times(M-1)$, where $M$ is the magnification factor of the system. The value of $M$ helps determine the spacing between beamlets on the kernel plane, setting the convolutional kernel size and maximum number without overlap.

A 2D array of spots is formed on the amplitude mask with spacing $sp_{K}=sp_{MLA}\times M$ and approximate size $M \times w_{LE} + sp_{MLA}$. These beamlets are individually modulated by different convolutional kernels on the mask with spacing $sp_{w}$ and total size $sz_{K}= sp_{w}k$. For experiments, we pad each convolutional kernel with a border of zeros to minimize optical crosstalk between different kernels. 

After the mask, light continues to propagate a distance $d_3=f \times M$ onto the corresponding photodetectors of the output channels. The photodetector spacing $sp_{PD} = sp_{w}d_{t}/d_3$ is given by relative distances between the different elements, where $d_{t}=d_1+d_2+d_3$. The distance between PD subarrays follows as above, with $sp_{PDA}= 2\times sp_{MLA}\times M$.

All light emitters for a single subarray are spaced such that their corresponding light spots map exactly onto the same convolutional kernels (Figure \ref{fig-3}b). In order for light from adjacent light emitters (LEs) in the same subarray to intersect the same kernels, they must pass through the centers of adjacent microlenses. This results in $sp_{MLA}/(f\times(M-1) = sp_{LE}/(f+f\times(M-1)$, simplifying to $sp_{MLA}/sp_{LE}=(M-1)/M$. Finally, the subarrays must have their relative spacing be a distance that is a multiple of $2 sp_{MLA} (2M)/(2M-1)$. This constraint is necessary to prevent kernel overlap.

After modulation by a convolutional kernel, the light from the adjacent LEs continues to propagate onto their corresponding PDs, diverging as they travel. For a convolution stride of 1 and a spacing $sp_{PD}=sp_{LE}$, this is satisfied when the distance from the LEs to the kernels is matched by the distance from the kernels to the PDs, or $d_1+d_2 = d_3$. Generally, different input and output spacings result in an overall scaling or shrinking of the input to output shapes.

Different input channels map onto the same output channel locations (Figure \ref{fig-3}c). Looking at light from LEs in adjacent input channels with the same relative pixel location, we find $sp_{LEA}(d_1+d_2)/d_{t} = sp_{off}$, where $sp_{off}$ represents the offset between kernels of neighboring channels. This combination of spacings results in the kernels on the amplitude mask spanning two different sets of offsets, illustrated in Figure \ref{fig-2}b.

\section{Optical convolution setup}

\begin{figure}[ht]
\centering
\includegraphics[width=11cm,clip]{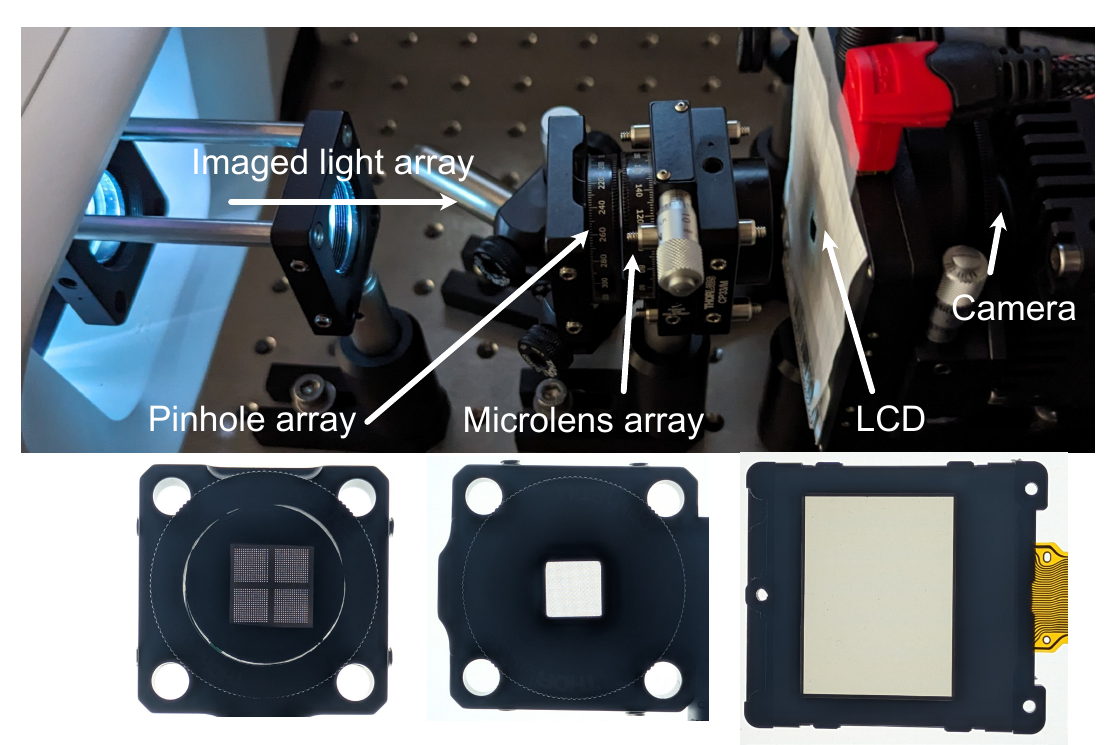}
\caption{Experimental setup. Light from a projector is imaged onto a photomask with pinholes precisely positioned to correspond with four input channels. The cropped light continues to propagate through a microlens array onto the kernel mask implemented using a liquid crystal display (LCD) and a pair of polarizers. The modulated light continues to propagate and is readout with a CMOS camera.}
\label{fig-4} 
\end{figure}

We demonstrate our multichannel optical convolution approach experimentally with four input channels and four output channels (Figure \ref{fig-4}). Here, we emulate a precisely positioned light emitter array required for our approach by imaging light from a projector (Viewsonic PG800D) onto a pinhole array. The pinhole array consists of four subarrays of $w_{LE} = 30\mu m$ square apertures with $sp_{LE} = 338\mu m$ and $sp_{LEA} = 6.35mm$ spacing between apertures on a film photomask (JD Photodata). By adhering a thin scattering layer (translucent tape) onto the illuminated side of the pinhole array and slightly defocusing the imaged projector output, we can control the brightness of individual pinholes by changing the corresponding pixel values on the projector. The scattering layer causes the light passing through the pinhole array to be more diffuse and have a more uniform emission profile.

This light continues to propagate through a microlens array with $sp_{MLA}=150\mu m$ (Thorlabs MLA150-5C) onto the convolution kernels. Convolutional kernels are encoded using an amplitude mask implemented with a liquid crystal display (Holoeye LC2012) and a pair of polarizers. The pixel size of the LCD is $36\mu m$, much less than $sp_{w}=169\mu m$, and allowing for both the precise placement of the convolutional kernels and the use of Gaussian shaped weights, which are used to reduce diffractive effects. The modulated light continues to propagate and is readout with a CMOS camera (FLIR Oryx-100G-71S7M).

\section{Simulation of optical convolution}
We first simulate the parameters of our optical setup using geometric raytracing. To estimate the expected optical performance of our approach, we used Monte-Carlo ray tracing to simulate the free-space optical propagation. $4$ input channels of size $8 \times 8$ were each convolved with $4$ kernels of size $3 \times 3$. No cropping or stride were used in this simulation, giving an output channel size of $10 \times 10$.

The simulation proceeds through a sequential geometric ray-tracing pipeline: we stochastically sample ballistic photons from each emitter using a uniform-angle emission model. The photons propagate to and through a microlens array via refraction, to an intermediate amplitude mask plane and finally to an output plane, where ray density histograms yield intensity images.

\begin{figure}[ht]
\centering
\includegraphics[width=10cm,clip]{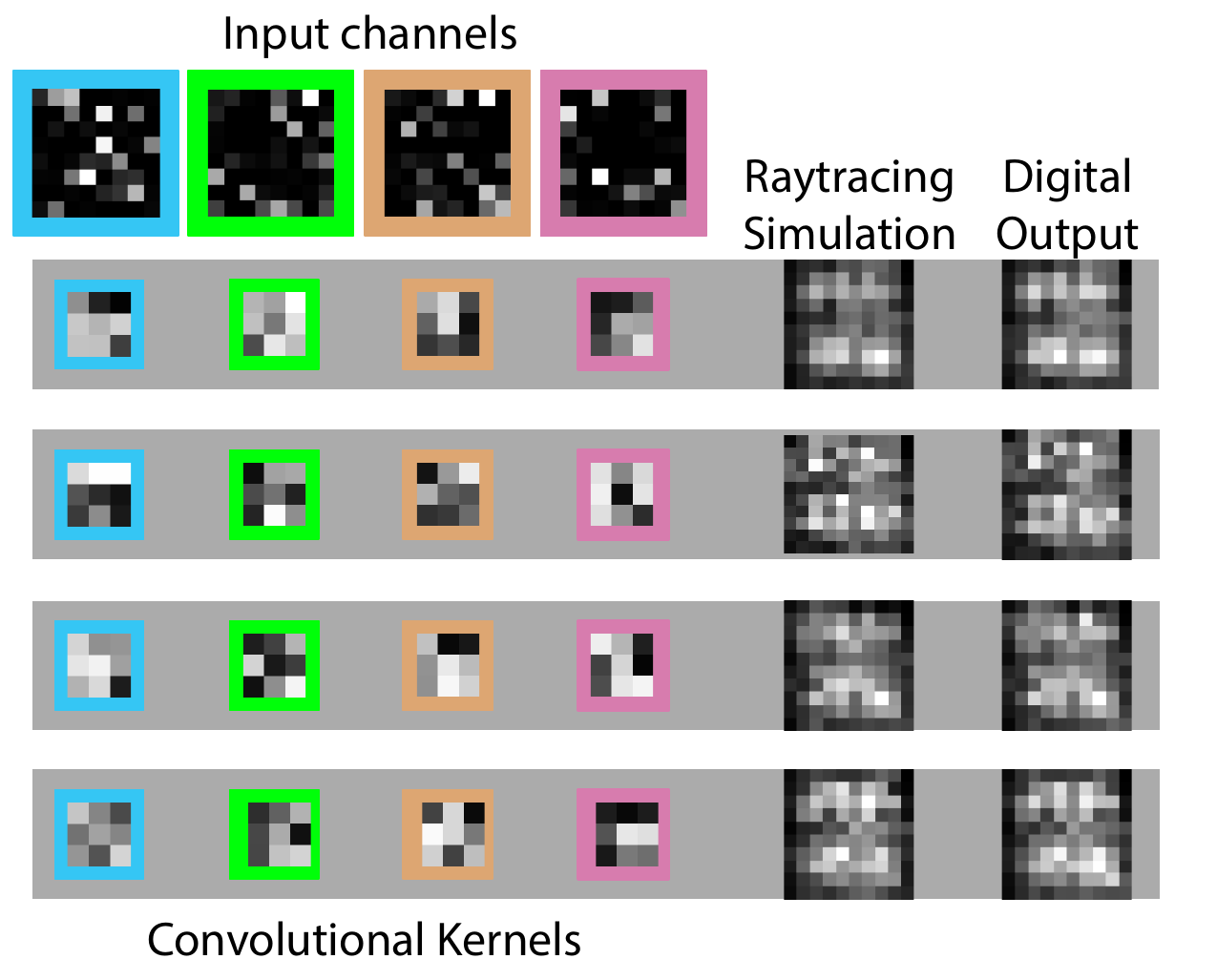}
\caption{Raytracing simulation of optical setup parameters. Photons are stochastically sampled from each emitter of each input channel (colors) and project through randomly generated convolution kernels onto an output plane where they are binned to the corresponding photodetector positions. Overall, the raytracing simulation results in an average Pearson’s correlation of $0.902$ for randomized input channels and convolutional weights.}
\label{fig-5} 
\end{figure}

Results from the simulation are depicted in Figure \ref{fig-5}. Input channels show a sparse, randomly distributed input convolved with a set of random $3 \times 3$ kernels. Optical raytracing produces a very similar result to ideal convolution values, with an overall Pearson's correlation of $0.902$. The reduction of in correlation is due to the angular distribution of light from the light emitters through different convolutional kernels.

\begin{figure}[ht]
\centering
\includegraphics[width=13cm,clip]{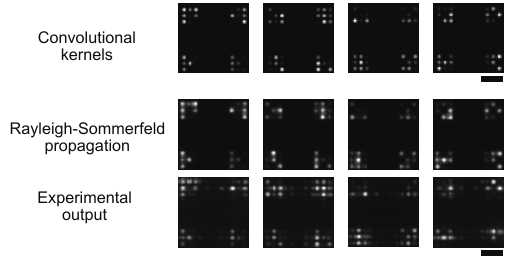}
\caption{Design of convolutional kernels and measurement of weights. Convolutional kernels corresponding to each of input channel (top row) are represented as apodised Gaussians with transmission amplitude corresponding to the weight value. These kernels are used to generate the simulated estimate (middle row) and to measure the experimental estimate (bottom row) of these weights after propagation through the optical system. Scale bar: $500\mu m$ (top), $1mm$ (middle, bottom)}
\label{fig-6} 
\end{figure}

The ray-tracing simulations do not account for the effects of the diffraction of light propagating through the optical setup, effects which ultimately limit the miniaturization of optical weights in our setup. To explore the effects of diffraction, we simulate the propagation of light using Rayleigh-Sommerfeld propagation from individual emitters through the system and compare the results with the measured outputs (Figure \ref{fig-6}).

For this simulation, we treat individual light emitters as independent sources with the emitter shape ($30\mu m$ square) and randomly initialized pixel phases between $0$ and $2\pi$. The same emitter is propagated multiple times with different input phase distributions that are each propagated using the Rayleigh-Sommerfeld diffraction integral through the optical setup and intensity averaged on the output plane. The microlens array is treated as a phase mask of a 2D array of lenslets with properties matching the experimental microlens array. The convolutional kernels are encoded as an amplitude mask composed of apodised Gaussians with amplitudes corresponding to the kernel weight. The apodised Gaussians are then binned into $36\mu m$ regions, corresponding with the pixel size of the LCD used.

The field continues to propagate onto the output plane where it is squared and binned into $100\mu m$ square regions, corresponding to the intensity projection onto photodetectors. These values are then compared with the design weights of the convolutional kernels. From these simulations including diffractive effects, we find a strong correspondence of the simulation output values to the designed kernel weights, with a Pearson’s correlation of 0.954.

We follow this up by measuring the convolutional weights through our experimental setup. Each light emitter is sequentially illuminated by the projector and propagates through the setup until the output intensities are recorded on the camera. The output values are binned and shifted to generate an estimate of the response of each convolutional kernel to a given light emitter. The mean of these responses is taken to produce the average kernel response of the optical setup. We find a good correspondence between the simulation output and the experimental measurements of the convolutional kernels with a Pearson’s correlation of 0.762.

\section{Multichannel convolution of handwritten digits}
\begin{figure}[ht]
\centering
\includegraphics[width=10cm,clip]{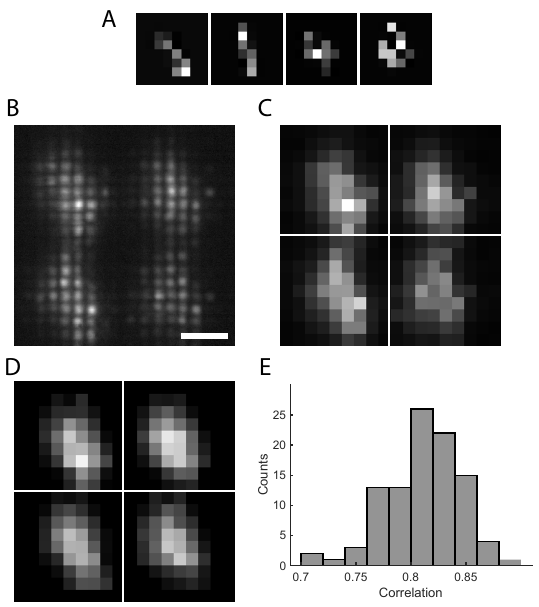}
\caption{Experimental demonstration of multichannel convolution with MNIST digits. (a) Example scaled-down MNIST digits used as four simultaneously presented input channels. (b) Output image on CMOS camera showing distribution of all output channels. Scale bar: $1mm$ (c) Extracted experimental convolutional output values. (d) Digital representation of convolution outputs. (e) Histogram of Pearson’s correlation values between experimental and digital convolution outputs.}
\label{fig-7} 
\end{figure}

We experimentally test one-shot multichannel convolution using handwritten digits \cite{deng2012mnist} to represent individual input channels (Figure \ref{fig-7}). The inputs are MNIST digits that have been bilinearly scaled down to an $8 \times 8$ image, corresponding to the input channel height and width. Each input channel is used to encode a separate digit and these inputs are simultaneously presented, propagating through the microlens array and convolutional kernels from Figure \ref{fig-6} onto the camera.

The camera output shows a $2 \times 2$ grid of output channels composed of well-isolated spots. Here, the different output channels are designed to not overlap over the designed input size. The center $100\mu m$ regions of these spots are binned and the binned values are extracted to the corresponding output channels. To assess the accuracy of our experimental approach, we calculate the correlation between the linearized measured outputs and digitally calculated values for each input set of four digits. We find good correspondence between the experimental and ground truth values with an average correlation of 0.82.

\section{Summary}
We have developed and experimentally demonstrated a multichannel free-space convolution approach. Our work shows for the first time an optical setup that can perform 2D spatial convolution from multiple input channels to multiple output channels in a single shot. This approach uses the geometric positioning of light emitters and convolutional kernels to develop a scalable, free-space system that requires only a single microlens array to route many optical channels. The kernels on the amplitude mask are shared, enabling efficient packing for large numbers of input and output channels.

Our proof-of-concept demonstration used a projector and a pinhole array to emulate the light emitters required for our approach. A more general and space-efficient implementation can be achieved with custom microLED or mutually incoherent VCSEL arrays, where emitters of specified size can be precisely positioned. Additionally, custom photodiode arrays would be better suited as photodetectors for this system. Using these approaches will enable our system to be scaled to very large matrix sizes. The geometry of this approach means kernel weights are efficiently packed and shared between inputs, meaning parallel computation from hundreds of channels is possible.

Overall, these results demonstrate the potential for optical approaches to perform the multichannel convolutional operations required by modern machine learning approaches. We believe this architecture highlights the power of spatial micro-organization in optical and optoelectronic components and these results can be expanded upon and used to improve the capability of existing and future optical coprocessors and accelerators.

\newpage
\bibliography{refs.bib}





\newpage

\linespread{1.5}

\newpage

\newpage

\newpage

\newpage

\newpage

\newpage

\newpage

\begin{table}[ht]
\centering
\renewcommand{\arraystretch}{1.3}
\fontsize{10pt}{10pt}\selectfont
\begin{tabular}{|c|c|c|c|}
\hline
\textbf{Parameter} & \textbf{Description} & \textbf{Relationship} & \textbf{Value} \\
\hline

$n$ & Input channel number &  & $4$ \\
$m$ & Output channel number &  & $4$ \\
$H$ & Input tensor height &  & $8$ \\
$W$ & Input tensor width &  & $8$ \\
$k$ & Kernel size &  & $3$ \\
\hline

$d_1$ & Distance from LE to MLA & $d_1 = f$ & $5.6\,\mathrm{mm}$ \\
$d_2$ & Distance from MLA to kernels & $d_2 = f\times(M-1)$ & $44.8\,\mathrm{mm}$ \\
$d_3$ & Distance from kernels to PDA & $d_3 = d_1+d_2 = f\times M$ & $50.4\,\mathrm{mm}$ \\
\hline

$M$ & System magnification & $M = d_3/d_1$ & $9$ \\
\hline

$w_{LE}$ & Width of LE & $M w_{LE} + sp_{MLA}> sz_{K}$ & $30\,\mu m$ \\
$sp_{LE}$ & Spacing between LE &  & $338\,\mu m$ \\
$sp_{LEA}$ & Spacing between LEAs & 
$sp_{LEA} \propto 2 sp_{MLA}\frac{2M}{2M-1}$ & $6.35\,\mathrm{mm}$ \\
\hline

$sp_{MLA}$ & Spacing between microlenses & 
${sp_{MLA}}/{sp_{LE}}=\frac{M-1}{M}$ & $150\,\mu m$ \\
$f$ & Focal length of microlens &  & $5.6\,\mathrm{mm}$ \\
\hline

$w_{w}$ & Width of kernel weight &  & $60\,\mu m$ \\
$sp_{w}$ & Spacing between kernel weights & $sp_{w} = sp_{PD}/2$ & $169\,\mu m$ \\
$sp_{K}$ & Kernel spacing (input channel) & $sp_{K}=M sp_{MLA}$ & $1.35\,\mathrm{mm}$ \\
$sp_{off}$ & Kernel spacing (output channel) & $sp_{off} = sp_{LEA}/2$ & $3.18\,\mathrm{mm}$ \\
$sz_{K}$ & Kernel size & $sz_{K} = sp_{w}k$ & $0.5\,\mathrm{mm}$ \\
\hline

$w_{PD}$ & Width of a PD &  & $100\,\mu m$ \\
$sp_{PD}$ & Spacing between PDs & $sp_{PD}=sp_{LE}$ & $338\,\mu m$ \\
$sp_{PDA}$ & Spacing between PDAs & 
$sp_{PDA}= 2 M sp_{MLA}$ & $2.7\,\mathrm{mm}$ \\
\hline

\end{tabular}
\caption{Optical parameters and their relationship for experimental setup, with a convolution stride of 1 and equal light emitter and photodetector spacing.}
\label{tab-1}
\end{table}

\end{document}